\documentclass[twocolumn,showpacs,preprintnumbers,amsmath,amssymb,nofootinbib]{revtex4}

\usepackage{graphicx}
\usepackage{epsfig}
\usepackage{color}

\newcommand{\beq}{\begin{equation}}
\newcommand{\eeq}{\end{equation}}
\newcommand{\bea}{\begin{eqnarray}}
\newcommand{\eea}{\end{eqnarray}}
\newcommand{\bwd}{\begin{widetext}}
\newcommand{\ewd}{\end{widetext}}

\begin{document}
\title{Generation of Coherent X-Ray Radiation Through Modulation Compression}

\author{Ji Qiang}
\email{jqiang@lbl.gov}
\affiliation{Lawrence Berkeley National Laboratory, Berkeley, CA 94720, USA}

\author{Juhao Wu}
\affiliation{SLAC National Accelerator Laboratory, Menlo Park, CA 94025, USA}

\date{\today}

\begin{abstract}
In this paper, we propose a scheme to generate tunable coherent X-ray radiation for future light source applications.
This scheme uses an energy chirped electron beam, a laser modulator, a laser
chirper and
two bunch compressors to generate a prebunched kilo-Ampere
current electron beam from a few tens Ampere electron beam out of a linac.
The initial modulation energy wavelength can be compressed by a factor of
$1+h_b R_{56}^a$ in phase space, where $h_b$ is the energy bunch length 
chirp introduced by the laser chirper, $R_{56}^a$ is the momentum compaction 
factor of the first bunch compressor.
As an illustration, we present an example to generate more than $400$ MW, $170$ attoseconds pulse,
$1$ nm coherent X-ray radiation using a $60$ Ampere electron beam
out of the linac and $200$ nm laser seed.
Both the final wavelength and the radiation pulse length in the proposed
scheme are tunable by
adjusting the compression factor and the laser parameters.

\end{abstract}

\maketitle


Coherent X-ray sources have important applications in
biology, chemistry, condensed matter physics, and material science.
In recent years, there is growing interest in generating single
attosecond X-ray radiation pulse using Free Electron Lasers
(FELs)~\cite{zholents1,zholents2,saldin2,wu2,zholents3,ding,xiang2,zholents4}.
In this paper we propose a scheme to generate density modulation at X-ray wavelength with application to producing single attosecond pulse coherent X-ray radiation.
This scheme compresses the initial seeded
energy modulation wavelength inside the electron beam by a factor of
$M$, $M=1+h_b R^a_{56}$. This compression factor can be made as
large as a few hundreds by using a laser chirper and
two bunch compressors to
reach the final short-wavelength modulation.
It differs from
other FEL seeding schemes
that make use of harmonics of the
seeded modulation wavelength to reach short-wavelength radiation~\cite{yu,timur,echo}, even though the harmonics of the compressed modulation in the proposed scheme
can still be used.
This makes the final radiation wavelength and
pulse length from the proposed scheme tunable by adjusting the compression factor and the laser parameters.
The proposed scheme also has significant advantages over our previously proposed compression scheme~\cite{qiang1}~\footnote{A similar scheme to Ref.~\cite{qiang1} using accelerator chirpers was also independently proposed by Ratner et al.~\cite{ratner,ratner2}.}, 
where the final modulation compression factor is $C$ instead of $M$. 
Here, $C$ is the compression factor from the first bunch compressor.
For a final compression factor of hundreds as proposed
in this paper, a small jitter of the initial beam energy chirp
will result in a large change of the compression factor.
In contrast to those previous schemes, this new scheme does not suffer from jitter largely as will be explained later. 
The new scheme also simplifies the hardware layout by using one fewer 
laser modulator and makes the whole scheme more compact than the previous
scheme.

\begin{figure}[tb]
   \centering
   \includegraphics*[angle=0,width=90mm]{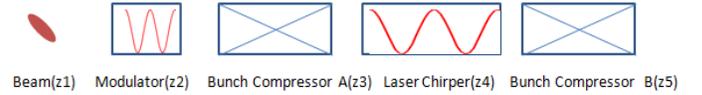}
   \caption{A schematic plot of the lattice layout of the modulation compression scheme.}
   \label{figxban3}
\end{figure}

A schematic plot of the modulation compression scheme is given
in Fig.~\ref{figxban3}.
It consists of an energy chirped electron beam, a seeding laser modulator, a bunch compressor A,
a laser chirper, and another bunch compressor B.
In this scheme, an energy chirped electron beam is sent into the seeding laser modulator to obtain initial energy modulation.
By doing the energy chirping first instead of energy modulation first, it 
avoids the potential distortion of the modulation from collective effects
during the process of chirping inside the linac.
This also allows one to use the full linac
to generate the initial energy-bunch length chirp. This is in contrast to most of the currently operating XFELs and those under design, where the residual energy chirp after the bunch compression has to be removed by running the electron bunch off crest after the bunch compressor. This makes the acceleration less effective. 
Inserting a chicane with a reasonable compression factor
inside such a linac will produce the same amount of chirp required in this compression scheme
and will reduce the total voltage needed from the linac by a factor of the compression factor.
For an electron with initial longitudinal coordinates $(z_0,\delta_0)$ before
chirping, where
$z$ is the relative longitudinal distance with respect to the reference particle,
and $\delta = \Delta E/E$ is the relative energy deviation, the
energy after the chirping and the first laser modulation will be:
\begin{eqnarray}
\delta_2 & = & \delta_1 +  A \frac{1}{1+\delta_1}
\frac{\sin(4N_w\delta_1\pi)}{4N_w\delta_1\pi}\sin(kz)\label{enemod}
\end{eqnarray}
where $\delta_1=\delta_0+h \ z$ represents the relative energy spread
after chirping from the linac, where $h$ is the energy chirp associated with the electron beam before the first bunch compressor, $A$ is the modulation amplitude
in the unit of the relative energy, $N_w$ is the number of periods of the
wiggler in the laser modulator, and $k$ is the wave number of the seeding laser.
Given the fact that the bunch length is normally below a millimeter,
the required initial chirp for the first bunch compressor is on the order
of tens m$^{-1}$, and the number of wiggler periods is small, 
then Eq.~(\ref{enemod}) can be
approximated as:
\begin{eqnarray}
\delta_2 & = & \delta_1 +  A \sin(kz).
\end{eqnarray}
Assume the  longitudinal phase space distribution 
after the laser energy modulation as
$F(z,(\delta-hz-A \sin(kz))/\sigma)$, 
after the beam transports through the bunch compressor A, the laser
chirper, and the bunch compressor B,
the final longitudinal phase
space distribution will be:
\begin{eqnarray}
f(z,\delta) & = & F\left(Mz-(R^b_{56}M+R^a_{56})\delta, \right.  \nonumber \\
& & [\delta(1+h R^a_{56} + (h_b + h M)R^b_{56})- (h_b+h M)z- \nonumber \\
& & \left. A \sin(kMz-k(R^b_{56}M+R^a_{56})\delta)]/\sigma \right)
\label{dist}
\end{eqnarray}
where 
\begin{eqnarray}
M & = & 1+h_b R^a_{56}
\label{mm}
\end{eqnarray}
denotes the total final modulation compression factor, 
$R^b_{56}$ is the momentum compaction factor of the second bunch compressor, and
$\sigma$ is the beam uncorrelated energy spread.
If the chirp $h_b$ introduced by the laser chirper 
is $-h C$, i.e. $h_b=-h C$, 
the total modulation compression factor $M$ will be
$C$, i.e. $M=C$.
If 
the beam is over unchirped inside the laser chirper, i.e. $h_b = (-h+\tilde{h})C$, the total modulation compression factor will be $M=(1+\tilde{h}R^a_{56})C$.
Here, ``over unchirp'' means not only to remove the original energy chirp
of the beam after the bunch compressor A but also to introduce another
energy chirp to the beam in the opposite direction to the original chirp.
This improves the final compression factor in the new scheme by a factor of $1 + \tilde{h} R^a_{56}$, where the $\tilde{h}$ stands for the extent of over unchirping of the initial chirp
inside the laser chirper.
It is also seen from Eq.~\ref{mm} that the final modulation compression factor
does not depend on the initial beam energy chirp jitter.
Meanwhile, using a lower compression factor ($\sim10$) from the first bunch compressor
while maintaining the same final compression factor ($\sim100$), 
the final beam current can also be made not sensitive to the beam energy chirp
jitter. 
By properly choosing 
the momentum compaction
factor of the second bunch compressor, i.e. 
\begin{eqnarray}
R^b_{56} = -R^a_{56}/M,
\label{rr}
\end{eqnarray}
the final longitudinal phase space distribution is reduced into:
\begin{eqnarray}
f(z,\delta) & = & F(Mz,[\delta - M \tilde{h}z- MA\sin(kMz)]/M \sigma)
\end{eqnarray}
The above distribution function represents a compressed modulation
in a chirped beam.
It should be noted that the momentum compaction factor of the second bunch compressor does not have to be opposite sign with respect to the first bunch compressor if the total compression factor $M$, i.e. 
$C$ or $1 + \tilde{h} R^a_{56}$, is made negative. 
In the above equations, we have also assumed a longitudinally frozen electron beam and a linear laser chirper instead of the actual sinusoidal function from the laser modulator. The effects of sinusoidal energy modulation from the laser chirper will be discussed below.
  
To generate a large energy chirping
amplitude in the chirper after the first bunch compressor is not
easily accessible
using a conventional RF linac structure. Using a laser modulator,
this chirp
can be easily achieved by taking advantage of the short wavelength and the high power of the laser.
The disadvantage of using such a laser chirper is that not the whole
electron beam will be uniformly unchirped due to the sinusoidal form
of the energy modulation. Only periodically separated
local segments of the beam will be correctly unchirped.
However, such a property can be used to
generate ultra-short coherent X-ray radiation by controlling the
fraction of the beam that can be properly unchirped using a few-cycle laser pulse with carrier envelope
phase stabilization.
In the following simulation, we will produce such an attosecond short-wavelength modulated beam as an illustration.

\begin{figure}[htb]
   \centering
   \includegraphics*[angle=0,width=70mm]{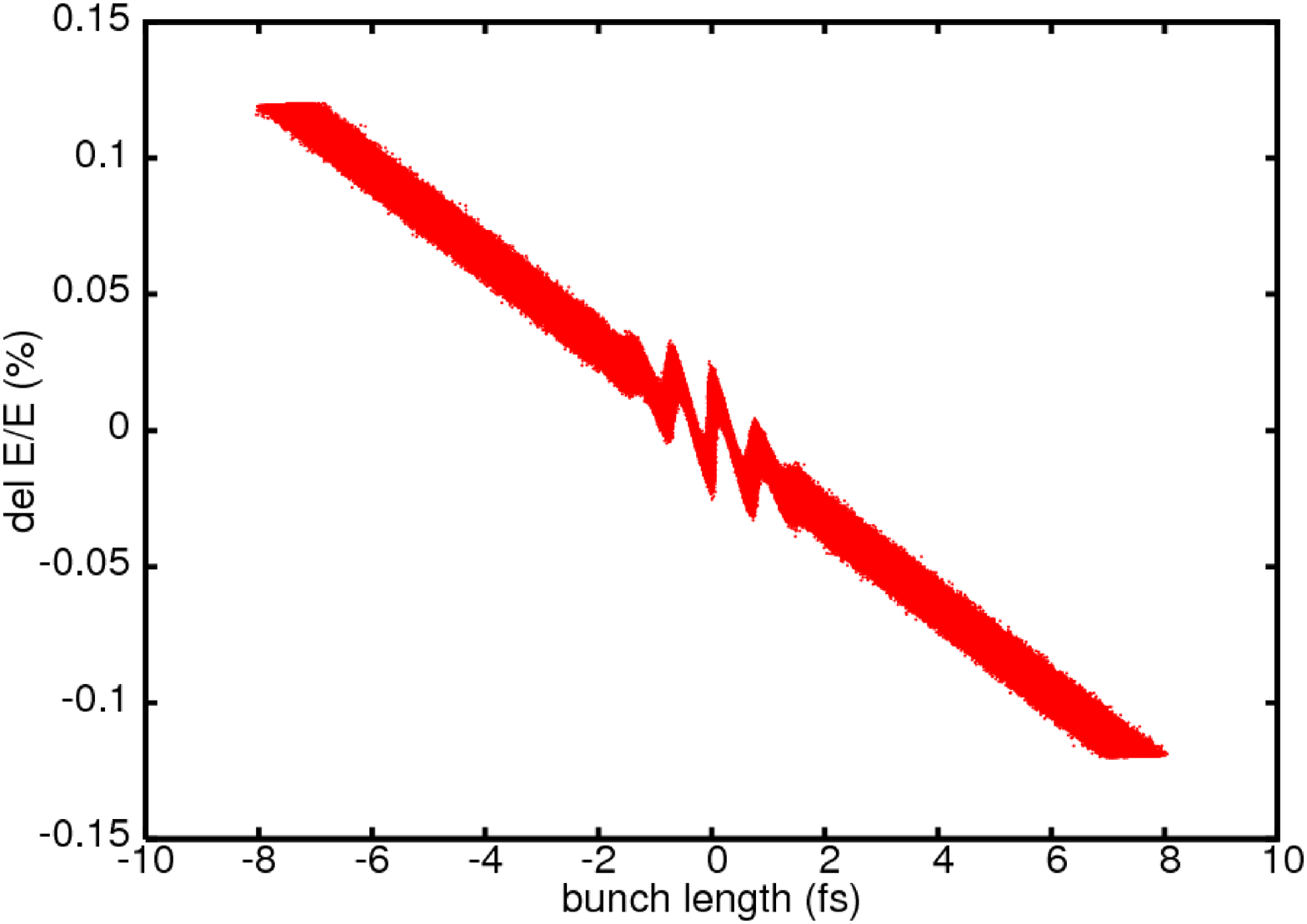} \\
   \includegraphics*[angle=270,width=70mm]{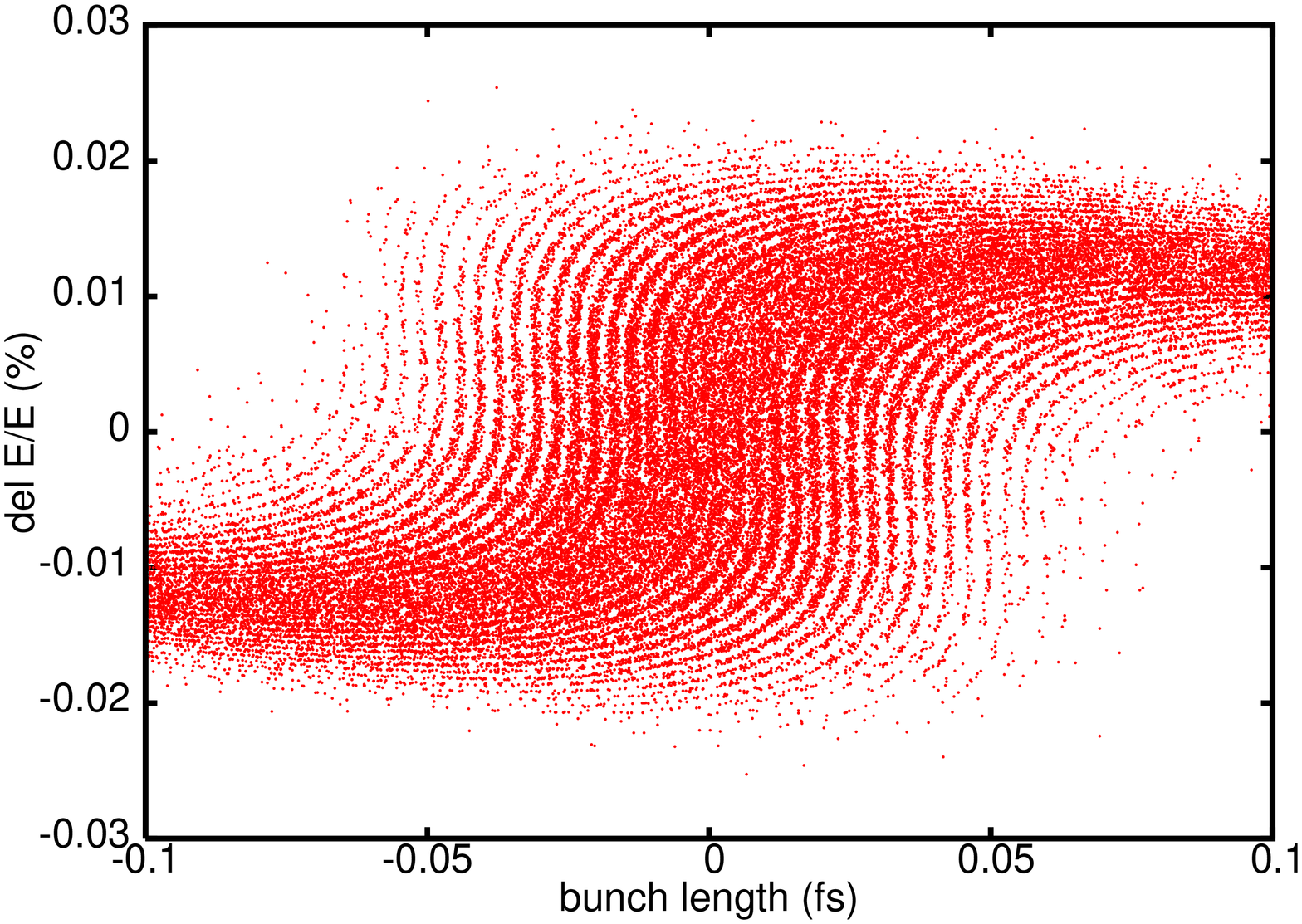}
   \caption{Longitudinal phase space after the compression scheme (top) and the zoom-in phase space around
the center of the beam (bottom).}
   \label{fig25a}
\end{figure}

\begin{figure}[htb]
   \centering
   \includegraphics*[angle=270,width=70mm]{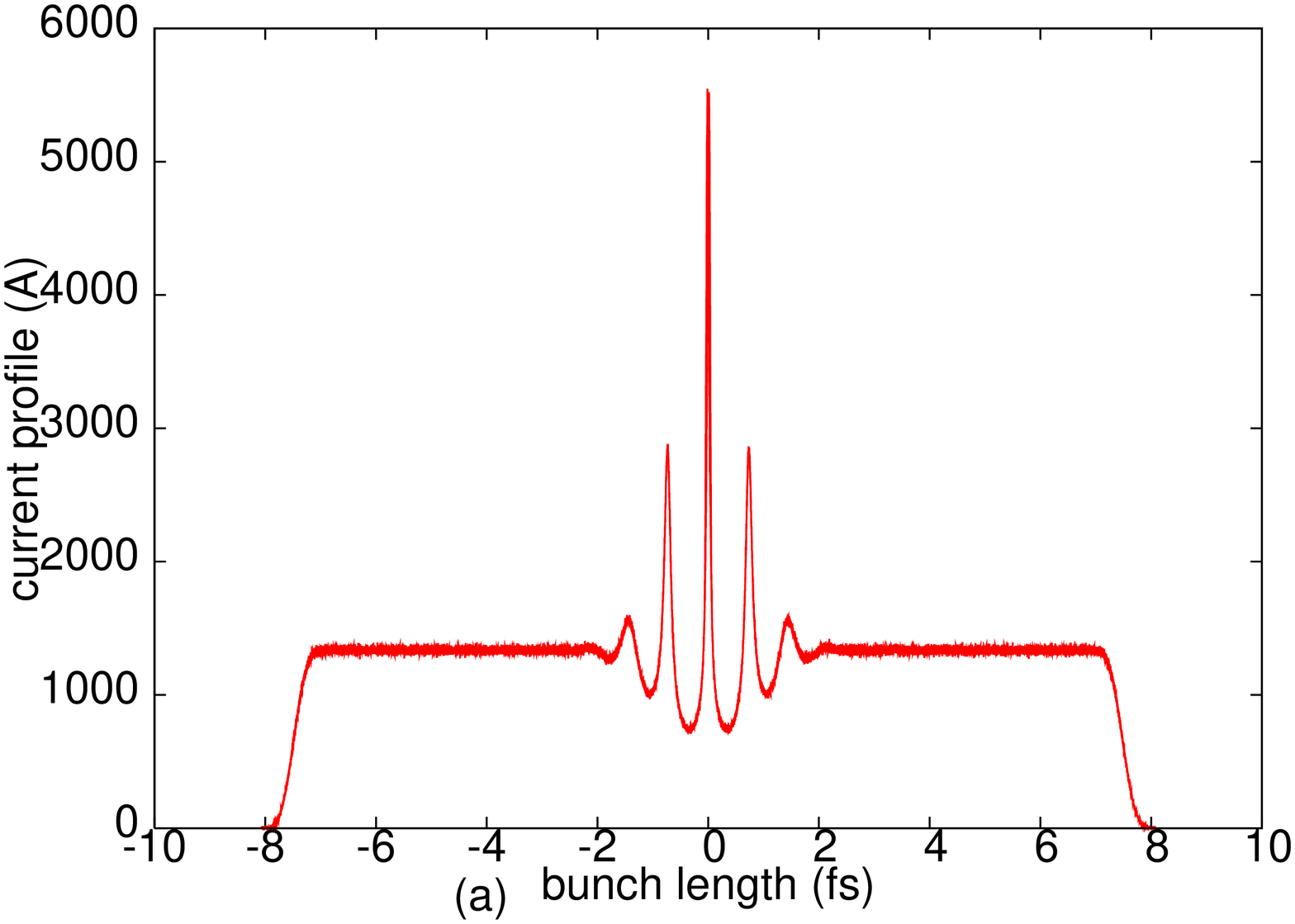}
   \includegraphics*[angle=270,width=70mm]{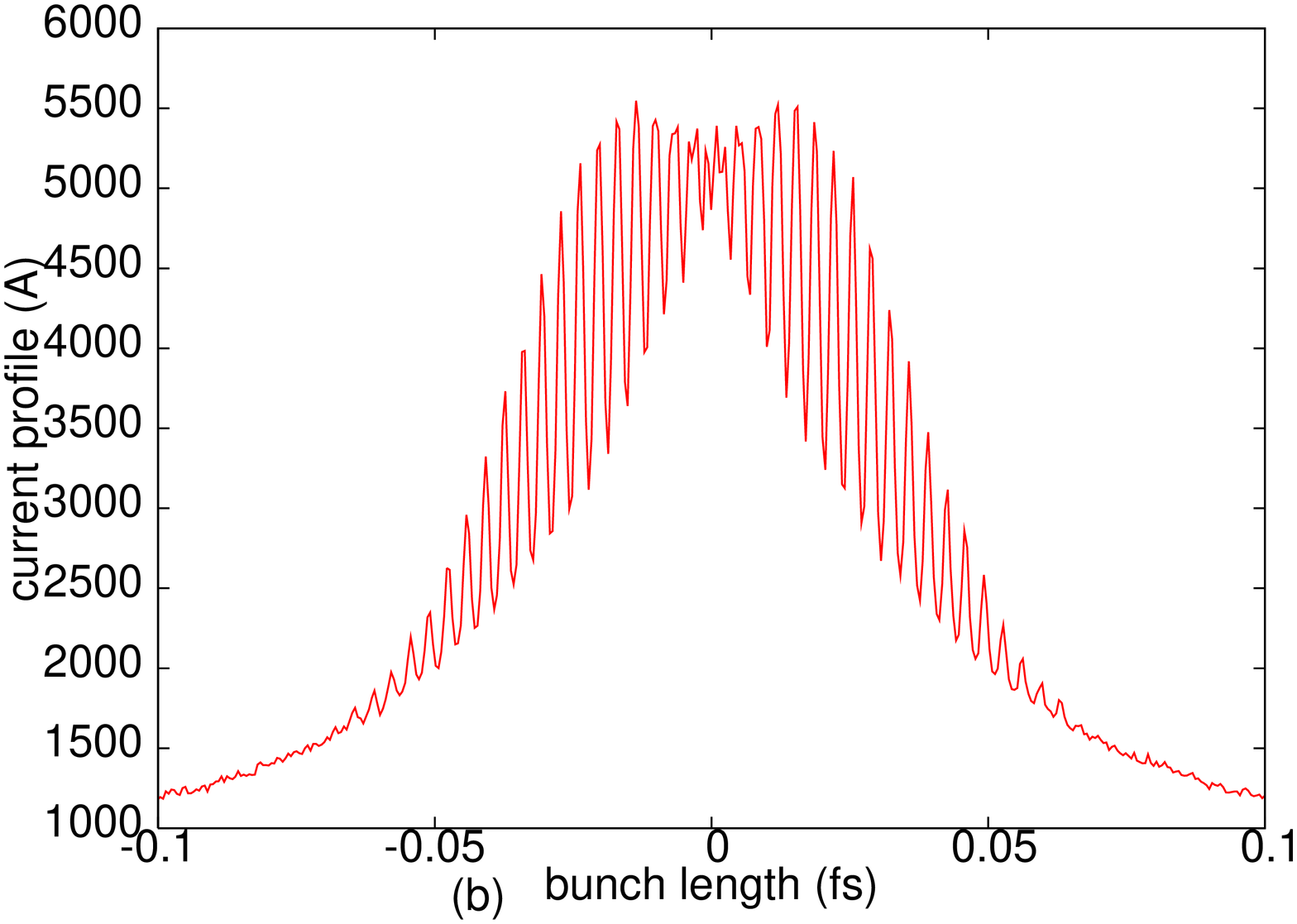}
   \caption{Beam current distribution after the compression scheme (a) and
the zoom-in current distribution around the center of the beam (b).}
   \label{figcur}
\end{figure}

The parameters of the electron beam and the lattice are summarized in
Table~\ref{pp}. 
\begin{table}[hbt]
   \centering
   \caption{Electron beam and lattice parameters in the illustration example}
   \begin{tabular}{lc}
       \toprule
           beam energy (GeV)        & 2                     \\
           beam current (A)      & 60                     \\
           initial enegy chirp (/m)      &     -23.95                 \\
           initial relative energy spread        &  $10^{-6}$                    \\
           laser modulation amplitude ($10^{-6}$)      & 1.2                    \\
           laser modulator wavelength (nm)      & 200                     \\
           laser modulator power (kW)      & 130                     \\
           $R_{56}$ of bunch compressor A  (cm)    & 4                     \\
           laser chirp amplitude ($10^{-4}$)      & 1.58                    \\
           laser chirp wavelength (nm)      & 200                     \\
           laser chirp power (GW)      & 2                     \\
           $R_{56}$ of bunch compressor B  (mm)    & -0.2                     \\
       \toprule
   \end{tabular}
\label{pp}
\end{table}
Here, a short uniform electron bunch ($100$ $\mu$m) with $20$ pC charge, $2$ GeV energy, $-23.95$ m$^{-1}$ energy-bunch length chirp, and an uncorrelated energy spread of $1 \times 10^{-6}$ is assumed at the beginning of the laser modulator.
The initial normalized modulation amplitude $A$ is $1.2 \times 10^{-6}$.
Assuming $1$ Tesla magnetic field in the wiggler with a total length of $33$ cm and a period of $11$ cm, this corresponds to about
$130$ kW $200$ nm wavelength
laser power.
After the modulator, we add an uncorrelated energy spread of $0.56$ keV to account for the synchrotron radiation effects inside the wiggler magnet using an
estimate from Refs.~\cite{zholents4,yy}.
After the initial seeding laser modulator,
the beam passes through a chicane bunch compressor.
Here, we have assumed that the $R_{56}$ of the chicane is about $4$ cm.
As the electron beam passes through a bending magnet, the quantum fluctuation of the
incoherent synchrotron radiation (ISR)
could cause the growth of the uncorrelated energy spread.
Such a growth of the uncorrelated energy spread might smear out the
modulation signal after the compression.
Here, we assume four $4$ meter long bending magnets inside the chicane each with $0.0655$ radian
bending angle in order to reduce the growth of the uncorrelated energy spread.
The rms energy spread induced by the ISR through such a chicane is about $0.44$ keV using an estimate from
Refs.~\cite{zholents4,chao}.
After the beam passes through this bunch compressor, the total bunch length of the beam is compressed down to
about $4$ $\mu$m due to a
factor of $25$ compression from the first bunch compressor.
This beam is transported through another few-cycle laser modulator with $200$ nm resonance wavelength.
This modulator works as an unchirper to remove the correlated energy chirp along the beam.
Here, the normalized amplitude of the laser is chosen as $1.58 \times 10^{-4}$ so that
the total modulation compression factor at the end of the second
bunch compressor is about $200$.
This modulation amplitude corresponds to about $2$ GW laser power using
the same wiggler as the first laser modulator.
After the beam transports through the laser modulator chirper, it passes through a dog-leg type bunch
compressor that can provide opposite sign $R^b_{56}$ compared to the chicane.
For a total compression factor of $200$ in this example,
the $R^b_{56}$ for the second bunch compressor is about $-0.2$ mm.
Figure~\ref{fig25a} shows the longitudinal phase space of the beam at the end of the second bunch compressor.
It is seen that after the second bunch compressor, only a small fraction of
beam is locally unchirped and further compressed. This also makes the above scheme
less sensitive to the time jitter between the laser and the electron beam.
The final energy spread in that zoom-in small section of the beam
is also small and good for generating coherent X-ray radiation
in an undulator radiator downstream.
Figure~\ref{figcur} shows the projected current profile for this prebunched beam.
Given the initial $60$ A current, the central prebunched current peak with $1$ nm wavelength modulation reaches
$5.5$ kA after the above compression scheme.
There are other lower current
spikes besides the central peak inside the beam. Those spikes will not
contribute significantly to the final $1$ nm attosecond radiation since
density microbunching in those spikes is very small due to
fact that the energy chirp at those locations does not produce
correct compression factor to satisty the Eq.~\ref{rr} of the
two bunch compressors. 
The length of the central prebunched beam is about a hundred attoseconds.
It is set by the half wavelength of the laser chirper.
Such a highly prebunched beam can be used to generate coherent attosecond X-ray radiation in a short undulator.

\begin{figure}[htb]
   \centering
   \includegraphics*[angle=270,width=70mm]{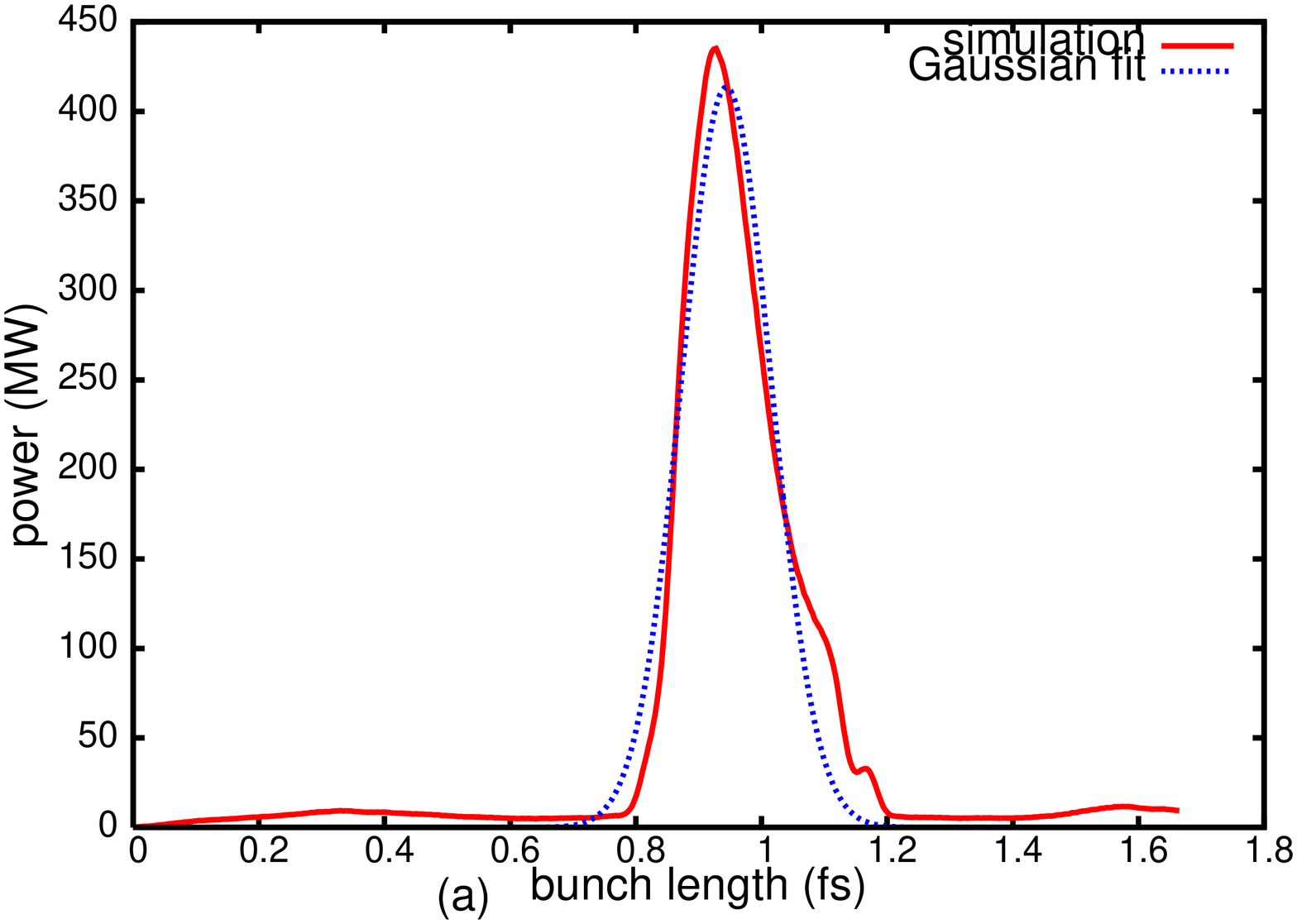}
   \includegraphics*[angle=270,width=70mm]{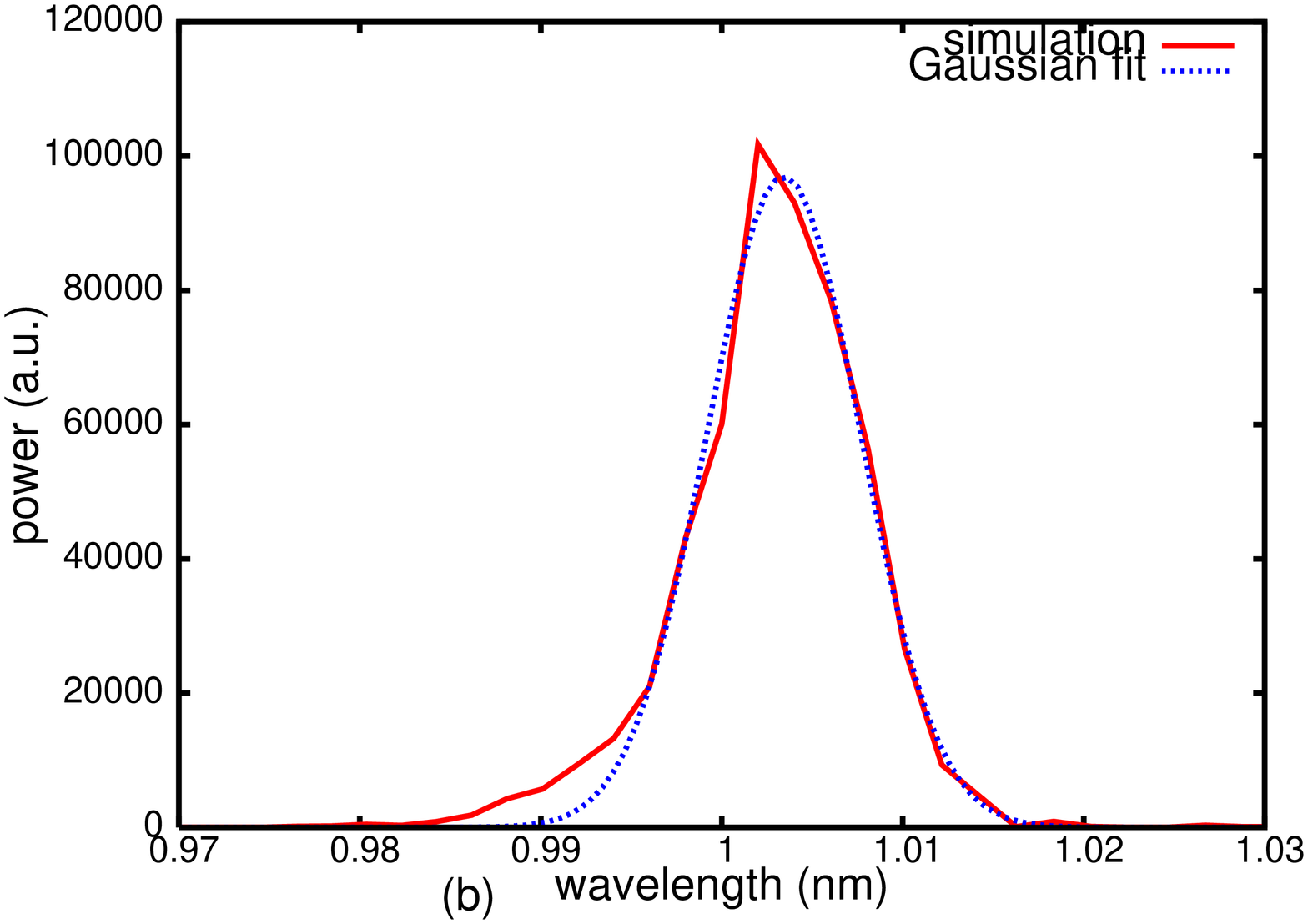}
   \caption{The radiation pulse temporal profile (a) at the end of the 1.5 m long undulator and the radiation pulse spectral profile (b) at the end of the 1.5 m long undulator.}
   \label{negR56sigpt1p2}
\end{figure}

To study the detailed radiation process and the radiation properties,
we use the {\small GENESIS} simulation code~\cite{Sven99} to calculate the coherent
X-ray radiation through a short undulator radiator. The normalized emittance
of the electron beam is chosen to be $0.2$ $\mu$m.
The length of the radiator is
$1.5$ m with an undulator period of $1.5$ cm.
The details of the radiation
properties at the end of the undulator are shown in Fig.~\ref{negR56sigpt1p2},
where
the radiation pulse power temporal profile is shown as the top subplot
and the radiation pulse spectral profile is shown as the bottom subplot.
The profiles are fit to Gaussian curves which give a temporal rms size of
$71.3$ as and a spectral rms width of $4.2$ pm.
This results in a phase space area of about $0.56$, which is close
to the transform limit of the X-ray pulse.
The full width at half maximum of the radiation pulse is about $170$ as,
which is larger than the width of the central modulated current density
distribution. This is due to
the slippage of the photon pulse with respect to the
electron bunch inside the radiator.

In this paper, we proposed a scheme to generate attosecond coherent
X-ray radiation through modulation
compression. This scheme allows one to tune the final X-ray radiation wavelength
by adjusting the compression factor. It also allows one to
control the final radiation pulse length by controlling
the laser chirper parameters and the initial compression parameters.
The proposed scheme requires a small initial seeding laser power due
to compression amplification.
This makes it possible to use the relatively low power ($\sim100$ kW)
and shorter wavelength laser ($\sim10$ nm) from the higher-order harmonic generation (HHG) scheme
to seed the electron beam
before the first bunch compressor~\cite{hhg}. This could result in a prebunched beam with a
modulation wavelength down to the hard X-ray regime using the above single stage
compression scheme.
The high compression factor in the proposed scheme
also lowers the required
initial current in order to achieve final kilo-Ampere electron beam current.
This helps maintain good beam
quality transporting through the linear accelerator.
In the above illustration example, a single $1$ nm attosecond pulse is generated
with a $200$ nm laser seeded $60$ Ampere current beam.
Another laser chirper, bunch compressor and radiator can be added after the
first X-ray radiator to generate a second attosecond pulse with different
radiation wavelength using another part of the beam to achieve two-color option~\cite{zholents4}.
Also, in this example, we assumed an opposite-sign second bunch compressor.
By tuning the laser chirper parameters, the momentum compaction factor of the
second bunch compressor can have the same sign as the first bunch compressor.
In this case, another chicane can be used in the above scheme to generate the attosecond coherent X-ray radiation as suggested in Reference~\cite{ratner} 
by using
an over compressed beam in the first bunch compressor.

\begin{acknowledgments}
We would like to thank Drs. J. Corlett, Y. Ding, B. Fawley, Z. Huang, R. Ryne,
R. Wilcox, D. Xiang, S. Zhang, and A. Zholents for useful discussions.
We would also like to thank Mr. W. Tabler for helping improve the readability of
the paper.
This research used computer resources at the National Energy Research
Scientific Computing Center and at the National Center for Computational
Sciences.
The work of JQ was supported by the U.S. Department of Energy under Contract No. DE-AC02-05CH11231 and the work of JW was supported by the U.S. Department of Energy under contract DE-AC02-76SF00515.
\end{acknowledgments}



\begin{thebibliography}{99} 

\bibitem{zholents1}
A.A. Zholents and W.M. Fawley, Phys. Rev. Lett. {\bf 92}, 224801 (2004).
\bibitem{zholents2}
A.A. Zholents and G. Penn, Phys. Rev. ST Accel. Beams {\bf 8}, 050704 (2005)
\bibitem{saldin2}
E.L. Saldin, E.A. Schneidmiller, and M.V. Yurkov, Phys. Rev. ST Accel. Beams {\bf 9}, 050702 (2006)
\bibitem{wu2}J. Wu, P.R. Bolton, J.B. Murphy, K. Wang, Optics Express {\bf 15}, 12749, (2007).
\bibitem{zholents3}A.A. Zholents and M.S. Zolotorev, New Journal of Physcis {\bf 10}, 025005 (2008).
\bibitem{ding}
Y. Ding, et al.,
Phys. Rev. ST Accel. Beams {\bf 12}, 060703 (2009).
\bibitem{xiang2}
D. Xiang, Z. Huang, and G. Stupakov, Phys. Rev. ST Accel. Beams {\bf 12}, 060701 (2009).
\bibitem{zholents4}
A.A. Zholents and G. Penn, Nucl. Instr. Meth. A {\bf 612}, 254 (2010).
\bibitem{yu}L. Yu, Phys. Rev. A {\bf 44}, 5178 (1991).
\bibitem{timur}T. Shaftan and L. Yu, Phys. Rev. E {\bf 71}, 046501 (2005).
\bibitem{echo}G. Stupakov, Phys. Rev. Lett {\bf 102}, 074801 (2009).
\bibitem{qiang1}
J. Qiang, Nucl. Instr. and Meth. A {\bf 621}, 39 (2010).
\bibitem{ratner}D. Ratner, Z. Huang, A. Chao, in proceedings of $31^{st}$ International Free Electron Laser Conference, Liverpool, UK, August 23-28, 2009, p. 200.
\bibitem{ratner2}D. Ratner, Z. Huang, A. Chao, Phys. Rev. ST Accel. Beams {\bf 14}, 020701 (2011).
\bibitem{yy}
E. Saldin, E. Schneidmiller, M. Yurkov, Nucl. Instr. and Meth. A {\bf 381},  545 (1996).
\bibitem{chao}
A.W. Chao, M. Tigner, {\it Handbook of Accelerator Physics and Engineering},
World Scientific, Singapore, 2006.
\bibitem{Sven99}S. Reiche, Nucl. Instr. and Meth. A {\bf 429}, 243 (1999).
\bibitem{hhg}E. J. Takahashi, et al., Appl Phys. Letts. {\bf 84}, p. 4 (2004).

\end{thebibliography}
\end{document}